\documentclass{elsart}

\usepackage{graphicx}
\usepackage{amssymb}

\begin{document}

\begin{frontmatter}

\title{Long-range memory stochastic model of the return in financial markets}

\author{V. Gontis\corauthref{cor1}},
\ead{gontis@itpa.lt}
\ead[url]{http://www.itpa.lt/~gontis}
\author{J. Ruseckas},
\author{A. Kononovi\v{c}ius}
\corauth[cor1]{Corresponding author.}
\address{Institute of Theoretical Physics and Astronomy of Vilnius
University, A.~Go\v{s}tauto 12, LT-01108 Vilnius, Lithuania }

\begin{abstract}
We present a nonlinear stochastic differential equation (SDE)
which mimics the probability density function (PDF) of the return
and the power spectrum of the absolute return in financial
markets. Absolute return as a measure of market volatility  is
considered in the proposed model as a long-range memory stochastic
variable. The SDE is obtained from the analogy with earlier
proposed model of trading activity in the financial markets and
generalized within the nonextensive statistical mechanics
framework. The proposed stochastic model generates time series of
the return with two power law statistics, i.e., the PDF and the
power spectral density, reproducing the empirical data for the one
minute trading return in the NYSE.
\end{abstract}

\end{frontmatter}

\section{Introduction}

High frequency time series of financial data exhibit sophisticated
statistical properties. What is the most striking is that many of
these anomalous properties appear to be universal. Vast amounts of
historical stock price data around the world have helped to
establish a variety of so-called stylized facts
\cite{Mantegna,Engle2,Plerou,Engle3,Podobnik,Bouchaud}, which can
be seen as statistical signatures of financial processes. The
findings as regards the PDF of the return and other financial
variables are successfully generalized within a non-extensive
statistical framework \cite{Tsallis}. The return has a
distribution that is very well fitted by $q$-Gaussians, only
slowly becoming Gaussian as the time scale approaches months,
years and longer time horizons. Another interesting statistic
which can be modeled within the nonextensive framework, is the
distribution of volumes, defined as the number of shares traded.

Interesting stochastic models related to the nonextensive
statistics include an ARCH process with random noise distributed
according to a $q$-Gaussian as well as some state-dependent
additive-multiplicative processes \cite{Queiros}. These models do
capture the distribution of returns, but not necessarily the
empirical temporal dynamics and correlations.
Additive-multiplicative stochastic models of the financial
mean-reverting processes provide a rich spectrum of shapes for the
probability distribution function (PDF) depending on the model
parameters \cite{Anteneodo}. Such stochastic processes model the
empirical PDF's of volatility, volume and price returns with
success when the appropriate fitting parameters are selected. Many
other fits are also  proposed, including exponential ones
\cite{Podobnik2} applicable for larger time scales.

Nevertheless, there is a necessity to select the most appropriate
stochastic models, able to describe volatility as well as other
variables in dynamical aspects and long-range correlation aspects.

There is empirical evidence that trading activity, trading volume,
and volatility are stochastic variables with the long-range
correlation \cite{Engle,Plerou2,Gabaix} and this key aspect is not
accounted for in some widely used models. The ARCH-like,
multiscale models of volatility, which assume that the volatility
is governed by the observed past price changes over different time
scales, have been recently proposed \cite{BorlandArx,QueirosEPL}.
Trading volume and trading activity are positively correlated with
market volatility. Moreover, trading volume and volatility show
the same type of long memory behavior \cite{Lobato}.

Recently we investigated analytically and numerically the
properties of stochastic multiplicative point processes
\cite{Gontis2004,KaulakysPRE}, derived a formula for the power
spectrum and related the model with the general form of the
multiplicative stochastic differential equation
\cite{KaulakysPhA,KaulakysJSTAT}. The extensive empirical analysis
of the financial market data, supporting the idea that the
long-range volatility correlations arise from trading activity,
provides valuable background for further development of the
long-ranged memory stochastic models \cite{Plerou2,Gabaix}. The
power law behavior of the autoregressive conditional duration
process \cite{SatoPRE} based on the random multiplicative process
and its special case the self-modulation process
\cite{TakayasuPhA}, exhibiting $1/f$ fluctuations, supported the
idea of stochastic modeling with a power law PDF and long memory.
A stochastic model of trading activity based on an SDE driven
Poisson-like process has been already presented in
\cite{Gontis2008}. We further develop an approach of modulating
the SDE with a closer connection to the nonextensive statistics in
order to model the dynamics of return in this paper.

Long memory (long-term dependence) has been defined in time domain
in terms of autocorrelation power law decay, or in frequency
domains in terms of power law growth of low frequency spectra.
Despite statistical methodology being developed for the data with
the long-range dependence and the solid mathematical foundations
of the area \cite{Beran}, let us consider behavior of the
financial variables only in the frequency domain, analyzing the
power spectral density.

In the second Section we present the nonlinear SDE generating a
signal with a $q$-Gaussian PDF and power law spectral density. In
the third Section we analyze the tick by tick empirical data for
trades on the NYSE for 24 shares and adjust the parameters of the
proposed equations to the empirical data. A short discussion and
conclusions are presented in the final section.

\section{The stochastic model with a $q$-Gaussian PDF and long memory}

Earlier we investigated stochastic processes with long-range
memory properties. Starting from the stochastic point process
model, which reproduced a variety of self-affine time series
exhibiting the power spectral density $S(f)\sim1/f^\beta$ scaling
as power $\beta$ of the frequency $f$ \cite{KaulakysPRE}, later we
introduced a Poisson-like process driven by the stochastic
differential equation. The latter served as an appropriate model
of trading activity in the financial markets \cite{Gontis2008}. In
this section we generalize an earlier proposed nonlinear SDE
within the nonextensive statistical mechanics framework to
reproduce the long-range memory statistics with a $q$-Gaussian
PDF. The $q$-Gaussian PDF of stochastic variable $r$ with variance
$\sigma_q^2$ can be written as
\begin{equation}
P(r)=A_q \exp_q\left(-\frac{r^2}{(3-q)\sigma_q^2}\right),
\label{eq:qGaussian}
\end{equation}
where $A_q$ is a constant of normalization and $q$ defines the
power law part of the distribution. $P(r)$ is introduced through
the variational principle applied to the generalized entropy
\cite{Queiros}
\[
S_{q}=k\frac{1-\int[p(r)]^q \mathrm{d} r}{1-q} .
\]
Here the $q$-exponential of variable $x$ is defined as
\begin{equation}
\exp_q(x)=(1+(1-q)x)^{\frac{1}{1-q}}
\end{equation}
and we assume that the $q$-mean $\mu_q=0$. With some
transformation of parameters $\sigma_q$ and $q$
\[
\lambda=\frac{2}{q-1}\,,\qquad r_0=\sigma_q\sqrt{\frac{3-q}{q-1}}
\]
we can rewrite the $q$-Gaussian in a more transparent form:
\begin{equation}
P_{r_0,\lambda}(r)=\frac{\Gamma(\lambda/2)}{\sqrt{\pi}r_0\Gamma(\lambda/2-1/2)}
\left(\frac{r_0^2}{r_0^2+r^2}\right)^{\frac{\lambda}{2}}.
\label{eq:qGaussian2}
\end{equation}
Looking for the appropriate form of the SDE we start from the
general case of a multiplicative equation in the Ito convention
with Wiener process $W$:
\begin{equation}
\mathrm{d} r=a(r)\mathrm{d} t+b(r)\mathrm{d} W . \label{eq:SDE1}
\end{equation}
If the stationary distribution of SDE (\ref{eq:SDE1}) is the
$q$-Gaussian (\ref{eq:qGaussian2}), then the coefficients of SDE
are related as follows \cite{Gar85}:
\begin{equation}
a(r) =
-\frac{\lambda}{2}\frac{r}{r_0^2+r^2}b(r)^2+b(r)\frac{\mathrm{d}
b(r)}{\mathrm{d} r} . \label{eq:ar}
\end{equation}
From our previous experience modeling one-over-f noise and trading
activity in financial markets \cite{Gontis2004,KaulakysPRE},
building nonlinear stochastic differential equations exhibiting
power law statistics \cite{KaulakysPhA,KaulakysJSTAT}, we know
that processes with power spectrum $S(f)\sim 1/f^{\beta}$ can be
obtained using the multiplicative term $b(r)\sim r^{\eta}$ or even
a slightly modified form $(r_0^2+r^2)^{\frac{\eta}{2}}$.
Therefore, we choose the term $b(r)$ as
\begin{equation}
b(r)=\sigma(r_0^2+r^2)^{\frac{\eta}{2}}
\end{equation}
and, consequently, by Eq.~(\ref{eq:ar}) we have the related
relaxation
\begin{equation}
a(r)=\sigma^2\left(\eta-\frac{\lambda}{2}\right)(r_0^2+r^2)^{\eta-1}r
.
\end{equation}
Then one gets the stochastic differential equation
\begin{equation}
\mathrm{d}
r=\sigma^2\left(\eta-\frac{\lambda}{2}\right)(r_0^2+r^2)^{\eta-1}r\mathrm{d}
t +\sigma(r_0^2+r^2)^{\frac{\eta}{2}}\mathrm{d} W.
\label{eq:simple}
\end{equation}
Note that  in the simple case $\eta=1$ Eq. (\ref{eq:simple})
coincides with the model presented in the article by Queiros
\emph{et al.} \cite{Queiros2} with
\begin{equation}
b(r) = \sqrt{\frac{\theta}{P(r)^{\frac{2}{\lambda}}}}\,,\qquad
a(r) = -\frac{\theta}{r_0^2}\left(\frac{\lambda}{2}-1\right)r
\end{equation}
We will investigate higher values of $\eta$ in order to cache
long-range memory properties of the absolute return in the
financial markets. We can scale our variables
\begin{equation}
x=\frac{r}{r_0}\,,\qquad t_{s} = \sigma^{2}r_0^{2(\eta-1)}t
\label{eq:scaling}
\end{equation}
to reduce the number of parameters and to get simplified
equations. Then SDE
\begin{equation}
\mathrm{d}
x=\left(\eta-\frac{\lambda}{2}\right)(1+x^2)^{\eta-1}x\mathrm{d}
t_s +(1+x^2)^{\frac{\eta}{2}}\mathrm{d} W_s \label{eq:power2}
\end{equation}
describes a stochastic process with a stationary $q$-Gaussian
distribution
\begin{equation}
P_{\lambda}(x)=\frac{1}{\sqrt{\pi}}\frac{\Gamma(\lambda/2)}{\Gamma(\lambda/2-1/2)}
\left(\frac{1}{1+x^2}\right)^{\frac{\lambda}{2}} \label{eq:PDF3}
\end{equation}
and the power spectral density of the signal $S(f)$
\begin{eqnarray}
S(f) & = & \frac{A}{f^{\beta}}\,,\qquad
\beta=1+\frac{\lambda-3}{2(\eta-1)}
\label{eq:spectrum1}\\
A & = &
\frac{(\lambda-1)\Gamma(\beta-1/2)}{2\sqrt{\pi}(\eta-1)\sin(\pi\beta/2)}
\left(\frac{2+\lambda-2\eta}{2\pi}\right)^{\beta-1}
\label{eq:spectrum}
\end{eqnarray}
with $0.5 < \beta < 2$, $4 - \eta < \lambda < 1 + 2\eta$ and $\eta
> 1$. Eqs. (\ref{eq:spectrum1}-\ref{eq:spectrum}) were first derived
for the multiplicative point process in
\cite{Gontis2004,KaulakysPRE} and generalized for the nonlinear
SDE (\ref{eq:simple}) in \cite{KaulakysPhA,KaulakysJSTAT}.
Although Eq. (\ref{eq:simple}) coincides with Eq. (15) in ref.
\cite{KaulakysJSTAT} only for high values of the variable
$r>>r_0$, these values are responsible for the power spectrum.
Note that the frequency $f$ in equation (\ref{eq:spectrum1}) is
the scaled frequency matching the scaled time $t_s$
(\ref{eq:scaling}). The scaled equations
(\ref{eq:scaling})-(\ref{eq:spectrum}) define a stochastic model
with two parameters $\lambda$ and $\eta$ responsible for the power
law behavior of the signal PDF and power spectrum. Numerical
calculations with Eq. (\ref{eq:power2}) confirm analytical
formulas (\ref{eq:PDF3}-\ref{eq:spectrum}) (see ref.
\cite{KaulakysJSTAT}).

We will need a more sophisticated version of the SDE to reproduce
a stochastic process with a fractured power spectrum of the
absolute return observable in financial markets. Having in mind
the statistics of the stochastic model (\ref{eq:power2}) defined
by Eqs. (\ref{eq:PDF3})-(\ref{eq:spectrum}) and numerical modeling
with more sophisticated versions of the SDE, we propose an
equation combining two powers of multiplicativity
\begin{equation}
\mathrm{d}
x=\left(\eta-\frac{\lambda}{2}-(x\epsilon^{\eta})^2\right)
\frac{(1+x^2)^{\eta-1}}{((1+x^2)^{\frac{1}{2}}\epsilon+1)^2}x\mathrm{d}
t_s
+\frac{(1+x^2)^{\frac{\eta}{2}}}{(1+x^2)^{\frac{1}{2}}\epsilon+1}\mathrm{d}
W_s. \label{eq:return2}
\end{equation}
Here $\epsilon$ divides area of $x$ diffusion  into two different
power law regions to ensure the spectral density of $|x|$ with two
power law exponents. A similar procedure has been introduced in
the model of trading activity \cite{Gontis2008}. The proposed new
form of the continuous stochastic differential equation enables us
to reproduce the main statistical properties of the return
observed in the financial markets. This provides an approach to
the market with behavior dependent on the level of activity and
exhibiting two stages: calm and excited. Equation
(\ref{eq:return2}) models the stochastic return x with two power
law statistics, i.e., the PDF and power spectral density,
reproducing the empirical power law exponents of the trading
return in the financial markets. At the same time, via the term
$(x \epsilon^{\eta})^2$ we introduce the exponential diffusion
restriction for the high values of $x$ as the markets in the
excited stage operate on the limit of nonstationarity. We solve
Eq.~(\ref{eq:return2}) numerically using the method of
discretization. Introducing the variable step of integration
\[
h_k=\kappa^2\frac{((x_k^{2}+1)^{\frac{1}{2}}\epsilon+1)^2}{
(x_k^{2}+1)^{\eta-1}}\,,
\]
the differential equation (\ref{eq:return2}) transforms to the
difference equation
\begin{eqnarray}
x_{k+1} & = & x_k+\kappa^2\left(\eta-\frac{\lambda}{2}-(x
\epsilon^{\eta})^2\right)x_k+\kappa(x_k^{2}+1)^{\frac{1}{2}}\varepsilon_k\\
t_{k+1} & = &
t_k+\kappa^2\frac{((x_k^{2}+1)^{\frac{1}{2}}\epsilon+1)^2}{
(x_k^{2}+1)^{\eta-1}} \label{eq:return3}
\end{eqnarray}
The continuous stochastic variable $x$ does not include any time
scale as the return defined in a time window $\tau$ should. Having
in mind that the return is an additive variable and depends on the
number of transactions in a similar way to trading activity, we
define the scaled return $X$ in the time period $\tau$ as the
integral of the continuous stochastic variable
$X=\int_{t}^{t+\tau}x(t_s)/\tau\,\mathrm{d} t_s$. Note that $\tau$
here is measured in scaled time units Eq. (\ref{eq:scaling}) and
will coincide with a one minute interval of empirical data. This
serves as an procedure of adjustment  to the real time scale for
scaled equations.

It is worth recalling that integration of the signal in the time
interval $\tau$ does not change the behavior of the power spectrum
for the frequencies $f<<\frac{1}{\tau}$. This is just the case we
are interested in for the long-range memory analysis of financial
variables and we can expect Eqs.
(\ref{eq:spectrum1}-\ref{eq:spectrum}) to work for the stochastic
variable $X$ as well. We analyzed the influence of signal
integration on the PDF in previous modeling of trading activity;
see ref. \cite{Gontis2004}. Integration of the nonlinear
stochastic signal increases the exponent of the power law tails in
the area of the highest values of the integrated signal. This
hides fractured behavior of the $X$ PDF, which arises for $x$ as a
consequence of the two powers in the multiplicative term of Eq.
(\ref{eq:return2}).

In Fig.~1 we demonstrate (a) the numerically calculated PDF of
$|X|$
 in comparison with the theoretical distribution $2P(x)$ Eq.~(\ref{eq:PDF3}) and
(b) the numerically calculated  power spectrum of $|X|$ with
parameters appropriate for reproducing statistics for the absolute
return in financial markets.
\begin{figure*}
\centering
\includegraphics[width=.45\textwidth]{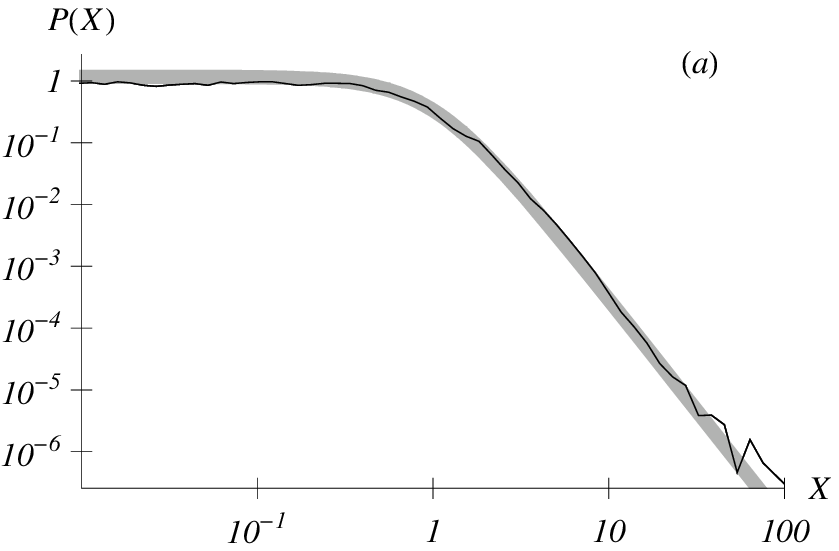}
\includegraphics[width=.45\textwidth]{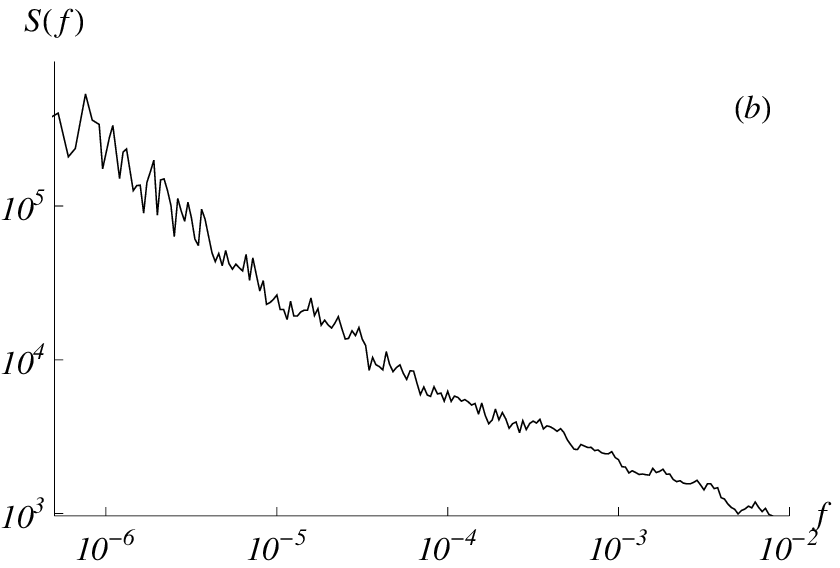}
\caption{(a) The numerically calculated PDF of
$|X|=|\int_{t}^{t+\tau}x(t)/\tau\,\mathrm{d} t_s|$ from
Eq.~(\ref{eq:return3}) (black thin line), in comparison with the
theoretical distribution $2P(x)$ Eq.~(\ref{eq:PDF3}) (gray thick
line), and (b) the numerically calculated power spectrum of $|X|$.
Parameters $\eta=5/2$, $\lambda=3.6$, $\tau=0.0001$ and
$\epsilon=0.01$ are selected to reproduce statistics for the
absolute return in financial markets.} \label{fig:1}
\end{figure*}

\section{Empirical analysis and model adjustment}

In this section we analyze the tick by tick trades of 24 stocks,
ABT, ADM, BMY, C, CVX, DOW, FNM, GE, GM, HD, IBM, JNJ, JPM, KO,
LLY, MMM, MO, MOT, MRK, SLE, PFE, T, WMT, XOM, traded on the NYSE
for 27 months from January, 2005, recorded in the Trades and
Quotes database. We sum empirical tick by tick returns into
one-minute returns to adjust the continuous stochastic model
presented. There is a problem in the use of a straightforward
procedure to determine $\eta$ from empirical data. One expects to
have $\eta\simeq 1$ when the return is assumed as a simple
stochastic variable \cite{Queiros2}. From our point of view the
straightforward SDE recovery procedures do not work, as the return
in real financial markets is at least double the stochastic
process influenced by long memory stochastic trading activity and
rapid price fluctuations. On the other hand, if one assumed $\eta=
1$, then long-range memory features of the process would be lost
\cite{KaulakysJSTAT}. Earlier we investigated the nonlinear
stochastic equations with $\eta\geq 3/2$, exhibiting the
long-range memory properties \cite{KaulakysPhA}, and proposed one
as an appropriate stochastic model of trading activity in the
financial markets \cite{Gontis2008}. Detailed analysis of the
empirical data from the NYSE provides evidence that long-range
memory properties of the return strongly depend on fluctuations of
trading activity. In Fig.~2 we demonstrate strong correlation of
the moving average of absolute returns per minute with the moving
average of trading activities (number of trades per minute). Here
for the empirical sequences of one-minute returns
$\left\{r_t\right\}_{t=1}^{T}$ or trading activities
$\left\{N_t\right\}_{t=1}^{T}$ we calculate moving averages
$\mathrm{MA}$ defined as the centered means for a selected number
of minutes $n$; for example, $\mathrm{MA}(r_t)$ is
\begin{equation}
\mathrm{MA}(r_t)=\frac{1}{n}\sum_{j=t-n/2}^{t+n/2-1}r_j.
\end{equation}
The best correlation can be achieved when the moving averages are
calculated in the period from $60$ to $100$ minutes.
\begin{figure*}
\centering
\includegraphics[width=.7\textwidth]{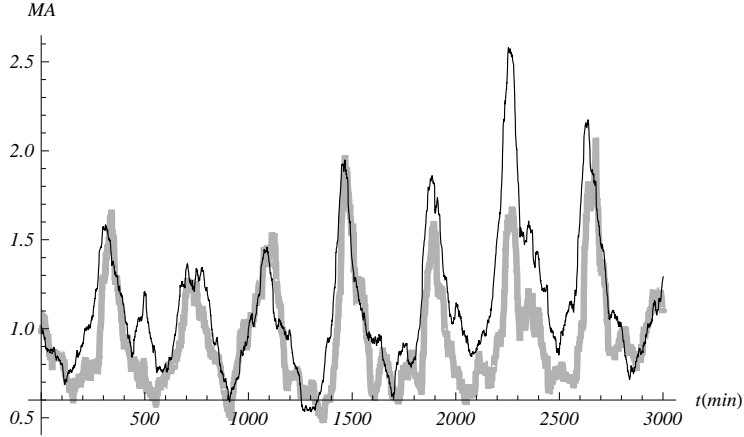}
\caption{An example of a moving average for 60 min  of empirical
absolute returns per minute (gray thick line) in comparison with
the corresponding moving average of trading activity, number of
trades per minute (black thin line). Scales are adjusted.}
\label{fig:2}
\end{figure*}

There are a lot of researchers investigating the power law
distribution of returns and trading activity  in the financial
markets \cite{BookDacorogna}. The $q$-Gaussian PDF is a reasonable
approximation to the empirical data \cite{Tsallis}. The power law
exponents for the extreme values of returns and trading activity
are nearly the same: $\lambda\simeq 4$ \cite{Gontis2008}.
Furthermore, fascinating statistical similarity of two financial
variables occurs in the power spectral density exhibiting
long-range memory properties with two scaling exponents
\cite{Gontis2008}. All these extraordinary sophisticated
statistical properties are reproducible using the SDE
(\ref{eq:return2}) introduced in the previous section.

Many non-equilibrium systems exhibit spatial or temporal
fluctuations of some parameter. There are two time scales: the
scale on which the dynamics is able to reach a stationary state,
and the scale for which the fluctuating parameter evolves. A
particular case is when the time needed for the system to reach
stationarity is much smaller than the scale at which the
fluctuating parameter changes. In the long term, the
non-equilibrium system is described by the superposition of
different local dynamics at different time intervals, which has
been called superstatistics \cite{beck-2003, abe-2007}.

In order to account for the  double stochastic nature of return
fluctuations - a hidden slowly diffusing long-range memory process
and rapid fluctuations of the instantaneous price changes --- we
decompose the empirical one-minute return series into two
processes: the background fluctuations and the high amplitude
rapid fluctuations dependent on the first one modulating. To
perform this decomposition we assume that the empirical return
$r_t$ can be written as instantaneous $q$-Gaussian fluctuations
with a slowly diffusing parameter $r_0$ dependent on the moving
average of the return $r_t$:
\begin{equation}
r=\xi\{r_0(\mathrm{MA}(r_t)),\lambda_2\}, \label{eq:MA}
\end{equation}
where $\xi\{r_0,\lambda_2\}$ is a $q$-Gaussian stochastic variable
with the PDF defined by Eq.~(\ref{eq:qGaussian2}) (the parameter
$q$ is $q=1+2/\lambda_2$). In Eq.~(\ref{eq:MA}) the parameter
$r_0$ depends on the modulating moving average of returns,
$\mathrm{MA}(r_t)$, and the empirically defined power law exponent
$\lambda_2$. From the empirical time series of the one-minute
returns $r_t$ one can draw histograms of $\xi$ corresponding to
defined values of the moving average $\mathrm{MA}(r_t)$. The
$q$-Gaussian PDF is a good approximation to these histograms and
the adjusted set of $r_0$ for selected values of
$\mathrm{MA}(r_t)$ gives an empirical definition of the function
\begin{equation}
r_0(\mathrm{MA}(r_t)) =
1+2.5\times|\mathrm{MA}(r_t)|.\label{eq:MA2}
\end{equation}
The $q$-Gaussians with $\lambda_2 = 5$ and linear function
$r_0(|\mathrm{MA}(r_t)|)$ (\ref{eq:MA2}) give a good approximation
of $\xi$ fluctuations for all stocks and values of modulating
$\mathrm{MA}(r_t)$. The long-term PDF of moving average
$\mathrm{MA}(r_t)$ can be approximated by a $q$-Gaussian with
$\bar{r}_0=0.2$ and $\lambda = 3.6$. All these empirically defined
parameters form the background for the stochastic model of the
return in the financial market.

\begin{figure*}
\centering
\includegraphics[width=.45\textwidth]{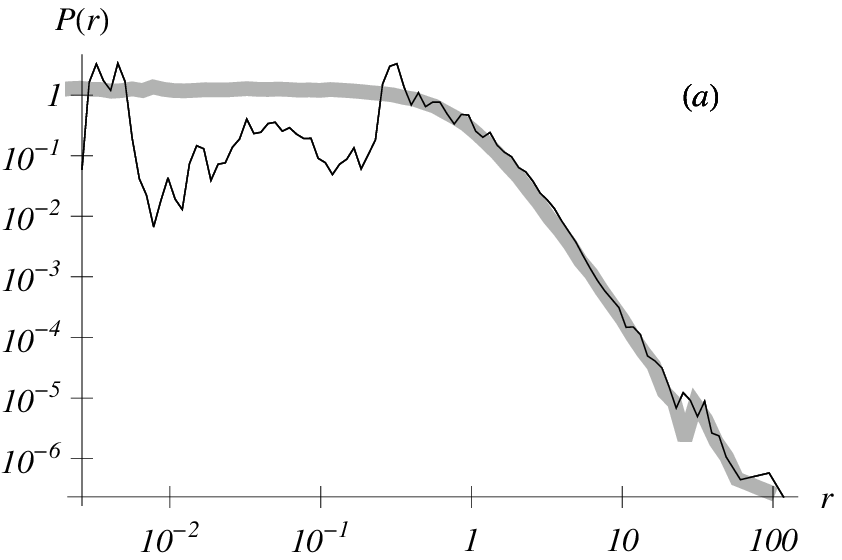}
\includegraphics[width=.45\textwidth]{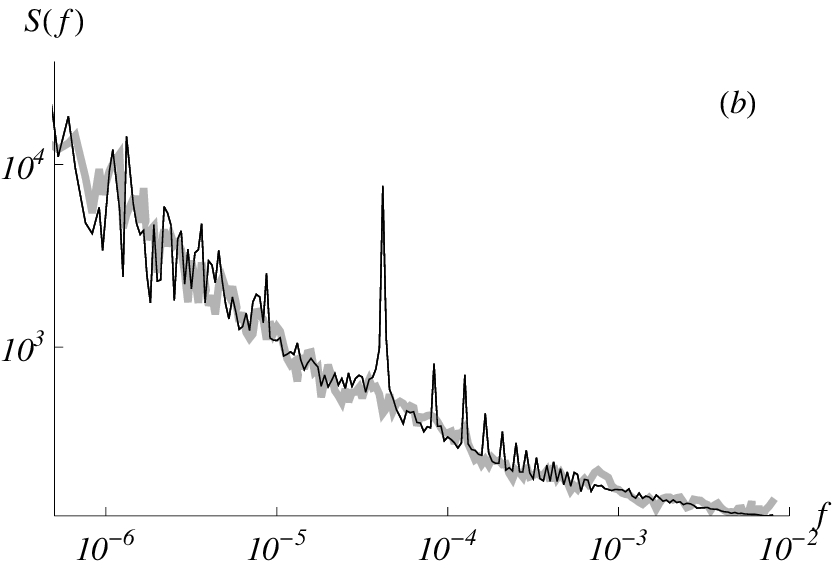}
\caption{Comparison of empirical and model
(\ref{eq:return2})-(\ref{eq:MA2}) statistics of one-minute returns
traded on  the NYSE, (a) the empirical (black thin line) and model
(gray thick line) PDF of normalized returns and (b) the empirical
power spectrum of normalized returns (black thin line) averaged
over 24 stocks and the model power spectrum (gray thick line)
averaged over 24 realizations. All parameters are as follows:
$\lambda_2=5$; $\bar{r}_0=0.2$; $\tau=0.0001/\sigma^2=60 s$;
$\lambda=3.6$; $\epsilon=0.01$ and $\eta=5/2$.} \label{fig:3}
\end{figure*}
We propose to model the long-range memory modulating stochastic
return $\mathrm{MA}(r_t)$ by $X = \bar{r}_0
\int_{t}^{t+\tau}x(t)/\tau\,\mathrm{d} t$, where $x$ is a
continuous stochastic variable defined by Eq.~(\ref{eq:return2}).
The remaining parameters $\epsilon$ and $\tau$ can be adjusted to
the empirical data and have values $\epsilon=0.01$ and
$\tau=0.0001/\sigma^2=60\,\mathrm{s}$. In Fig.~3 we provide a
comparison of the empirical PDF, averaged over 24 stocks of NYSE,
of one-minute returns normalized to the standard deviation and the
power spectrum with the corresponding statistics of the proposed
double stochastic model. This serves as an evidence of possibility
of modeling the financial variables using nonlinear stochastic
equations with elements of nonextensive statistics. Noticeable
difference in theoretical and empirical PDFs for small values of
$X$ are related with the prevailing prices of trades expressed in
integer values of cents. Obviously we do not account for this
discreteness in our continuous description. In the empirical power
spectrum one-day resonance --- the largest spike with higher
harmonics --- is present. This seasonality --- an intraday
activity pattern of the signal --- is not included in the model
either and this leads to the explicable difference from observed
power spectrum.

\section{Discussion and conclusions}

In the previous work \cite{Gontis2008} we provided evidence that
long-range memory fluctuations of trading activity in the
financial markets may be considered as the background stochastic
process responsible for the fractal properties of other financial
variables. This background stochastic process can be reproduced
using a nonlinear SDE (\ref{eq:return2}) with multiplicative noise
composed of two powers of a stochastic variable. The two powers in
the SDE reveal different behaviors of the market in the periods of
different trading activity. In this paper we generalized the form
of the background SDE within the nonextensive statistical
mechanics framework to reproduce fascinating statistical
properties of the financial variables with a $q$-Gaussian PDF and
fractured behavior of the power spectrum.

In the prevailing relatively calm periods, with $x<1/\epsilon$ and
multiplicativity specified by $\eta=5/2$, markets behave as
stationary stochastic processes with a $q$-Gaussian PDF, $q\sim
1+2/5=1.4$. In the periods of excited behavior, when
$x>1/\epsilon$, the PDF approaches a nonstationary regime,
$\lambda\sim 2$. This leads to the excess values of financial
variables, which have to be restricted by the additional limits in
the SDE, term $(x \epsilon^{\eta})^2$ giving the exponential
restriction of diffusion at the excess value $x\sim 1/
\epsilon^{\eta}$. These rare escapes of a continuous stochastic
variable smoothed by an integration procedure do not very
considerably contribute to the main PDF of a financial variable.
However, these escapes condition the behavior of the power
spectrum, reducing the exponent $\beta$ of the power law
distribution $S(f)\sim 1/f^\beta$ in the region of higher
frequencies. In the case of the return, the background stochastic
process defined by Eq.~(\ref{eq:return2}) is hidden by the
secondary high amplitude $q$-Gaussian stochastic process
$\xi\{r_0,\lambda_2\}$. Though the background fluctuations are
considerably lower than the secondary ones, this drives the whole
process through the empirically defined Eqs.~(\ref{eq:MA}) and
(\ref{eq:MA2}).

The generalized new form of the continuous stochastic
differential, equation (\ref{eq:return2}), enables us to reproduce
the main statistical properties of the return, observed in the
financial markets. All parameters introduced are recoverable from
the empirical data and are responsible for the specific
statistical features of real markets. The model does capture the
distribution of the return, the empirical temporal dynamics and
correlations evaluated through the power spectral density of
absolute return. The model definition with two powers of
multiplicative noise enables us to reproduce the power spectral
density with two different scaling exponents, as observed in the
empirical data. Stochastic modeling of the financial variables
with the nonlinear SDE is consistent with the nonextensive
statistical mechanics and provides new opportunities to capture
empirical statistics in detail.

\section*{Acknowledgements}
We would like to express our appreciation to B. Kaulakys for his
valuable advice and remarks. The authors acknowledge the support
by the Agency for International Science and Technology Development
Programs in Lithuania and EU COST Action MP0801 "Physics of
Competition and Conflicts".


\begin{thebibliography}{23}

\bibitem{Mantegna} R.N. Mantegna H.E. Stanley, Nature 376 (1995)
46-49.
\bibitem{Engle2} R. Engle, Econometrica 66 (1998) 1127-1162.

\bibitem{Plerou} V. Plerou, P. Gopikrishnan, L.A. Amaral, M. Meyer, H.E. Stanley,
Phys. Rev. E 60 (1999) 6519-6529.

\bibitem{Engle3} R. Engle,  Econometrica 68 (2000) 1-22.

\bibitem{Podobnik} P.Ch. Ivanov, A. Yuen, B. Podobnik,
Y. Lee, Phys. Rev. E 69 (2004) 056107.

\bibitem{Bouchaud} J.B. Bouchaud, M. Potters,
 \textit{Theory of Financial Risks and Derivative Pricing},
Cambridge University Press, Cambridge, 2004.

\bibitem{Tsallis} Cf.M. Gell-Mann, C. Tsallis, \textit{Nonextensive Entropy -
Interdisciplinary Applications}, Oxford University Press, NY,
2004.

\bibitem{Queiros} S.M. Duarte Queiros, C. Anteneodo, C. Tsallis, \textit{Power-law
distributions in economics: a nonextensive statistical approach},
in: Proc. Of SPIE 5848 (2005) 151. physics/0503024.

\bibitem{Anteneodo} C. Anteneodo, R. Riera, Phys. Rev. E 72 (2005)
026106.

\bibitem{Podobnik2} B. Podobnik, D. Horvatic, A. Petersen H.E.
Stanley, Europhysics Letters 85 (2009) 50003.


\bibitem{Engle} R.F. Engle, A.J. Paton, Quant. Finance 1
(2001) 237.

\bibitem{Plerou2} V. Plerou, P. Gopikrishnan, X. Gabaix et al, Quant.
Finance 1 (2001) 262.

\bibitem{Gabaix} X. Gabaix, P. Gopikrishnan, V. Plerou,  H.E. Stanley,
Nature 423 (2003) 267.

\bibitem{BorlandArx} L. Borland, \textit{On a multi-timescale statistical feedback model for volatility
 fluctuations}, 2004 arXiv:cond-mat/0412526.

\bibitem{QueirosEPL} S.M. Duarte Queiros, EPL, 80 (2007) 30005.

\bibitem{Lobato} I.N. Lobato, C. Velasco, J. Bus. Econom. Statist. 18 (2000)
410-427.

\bibitem{Gontis2004} V. Gontis, B. Kaulakys, Physica
A 343 (2004) 505-514.

\bibitem{KaulakysPRE} B. Kaulakys, V. Gontis, M. Alaburda, Phys. Rev. E
71 (2005)  051105.

\bibitem{KaulakysPhA} B. Kaulakys, J. Ruseckas, V. Gontis, M. Alaburda, Physica A 365 (2006)
217.

\bibitem{KaulakysJSTAT} B. Kaulakys, M. Alaburda, J. Stat. Mech.
(2009) P02051.

\bibitem{SatoPRE} A-H. Sato, Phys. Rev. E
69 2004 047101.

\bibitem{TakayasuPhA} M. Takayasu, H. Takayasu, Physica A
324 (2003)  101.

\bibitem{Gontis2008} V. Gontis, B. Kaulakys, J. Ruseckas, Physica
A 387 (2008) 3891-3896.

\bibitem{Gar85}  C.W. Gardiner,  \textit{Handbook of Stochastic Methods for Physics, Chemistry and Natural Sciences}, Springer-Verlag, Berlin, 1985.

\bibitem{Beran} J. Beran, \textit{Statistics for long-Memory Processes}, Chapman \&
Hall, New York, 1994.

\bibitem{Queiros2} S.M. Duarte Queiros, L.G. Moyano, J. de Souza,
C. Tsallis, Eur. Phys. J. B 55 (2007)  161-167.

\bibitem{BookDacorogna} M.M. Dacorogna, R. Gencay, U.A. Muller,
R.B. Olsen, O.V. Pictet, \textit{An Introduction to High-Frequency
Finance}, Academic Press, San Diego, 2001.

\bibitem{beck-2003} C. Beck and E. G. Cohen, Physica A 322 (2003) 267.

\bibitem{abe-2007} S. Abe, C. Beck, and E. G. D. Cohen, Phys.Rev.E 76 (2007)
031102.


\end{thebibliography}
\end{document}